\documentclass[aps,prb,twocolumn
,showpacs]{revtex4}

\usepackage{amsmath,amssymb}
\usepackage{graphicx}
\usepackage{dcolumn}
\usepackage{bm}

\def\rnum#1{\expandafter{%
\romannumeral #1}}
\def\Rnum#1{\uppercase\expandafter{%
\romannumeral #1}}

\begin{document}


\title{Vector chirality and inhomogeneous magnetization 
in frustrated spin tubes in high magnetic fields}

\author{Masahiro Sato and T\^oru Sakai}
\affiliation{Synchrotron Radiation Research Center, 
Japan Atomic Energy Agency, Sayo, Hyogo 679-5148, Japan and CREST JST, Japan}



\date{\today}

\begin{abstract}
The low-energy physics of 
three-leg frustrated antiferromagnetic spin-$S$ tubes 
in the vicinity of the upper critical field are studied. 
Utilizing the effective field theory based on the spin-wave approximation, 
we argue that in the intermediate-interchain-coupling regime, 
the ground state exhibits a vector chiral order or an inhomogeneous 
magnetization for the interchain (rung) direction and the low-energy 
excitations are described by a one-component Tomonaga-Luttinger liquid 
(TLL). 
In both chiral and inhomogeneous phases, 
the $Z_2$ parity symmetry along the rung direction 
is spontaneously broken. 
It is also predicted that 
a two-component TLL appears and all the symmetries are restored 
in the strong-rung-coupling case. 
\end{abstract}


\maketitle

\section{Introduction}
Frustrated spin systems~\cite{frust} 
have been continuously explored for more than five decades. 
Frustration is 
considered as an important keyword to generate exotic, 
unconventional magnetic orders, disorders and excitations including 
even spin-liquid states. 
Actually, frustrated systems have provided several peculiar 
concepts and phenomena so far: resonating-valence-bond picture, 
noncollinear orders, 
symmetry-unrelated degeneracy, order-by-disorder mechanism, etc.


In recent years, frustrated magnets
containing four-spin exchanges as well as 
standard two-spin ones have been intensively 
studied.~\cite{4spin} 
In such magnets, fascinating magnetic orders 
(nematic, chiral, dimer orders, etc.), 
which order parameter is defined by products of spin operators, 
are shown to be present. In a sense, 
these new orders are a natural consequence of the four-spin
exchange because for such an interaction, 
it is possible to perform a mean-field approximation, 
$S_i^\alpha S_j^\beta S_k^\gamma S_l^\delta\to
\langle S_i^\alpha S_j^\beta \rangle  S_k^\gamma S_l^\delta+
S_i^\alpha S_j^\beta \langle S_k^\gamma S_l^\delta \rangle 
- \langle S_i^\alpha S_j^\beta \rangle 
\langle S_k^\gamma S_l^\delta \rangle$.    
Furthermore, it is well known that 
effects of four-spin exchanges are fairly small 
in a large number of real magnets. Thus, to discover 
intriguing magnetic orders within spin systems 
{\it containing only two-spin exchanges} 
could more stimulate many experimentalists and 
would be theoretically a more challenging issue.

In one dimension, as representatives of 
geometrically frustrated spin systems with only two-spin exchanges, 
one can consider zigzag spin chains and three-leg antiferromagnetic (AF)
spin tubes, i.e., ladders with a periodic boundary condition (PBC)
along the interchain (rung) direction. 
In this paper, we study the latter model in a
magnetic field. The Hamiltonian is written as 
\begin{eqnarray}
\label{tube}
{\cal H} = \sum_{l=1}^3\sum_{j}\left[J\vec S_{l,j}\cdot\vec S_{l,j+1}
+J_\perp\vec S_{l,j}\cdot\vec S_{l+1,j}
-HS_{l,j}^z\right],
\end{eqnarray}
where $\vec S_{l,j}$ is spin-$S$ operator on site $j$ of the $l$th chain 
($l=1,2,3$), $J>0$ ($J_\perp>0$) is the intrachain (interchain)
coupling, and the PBC $\vec S_{4,j}=\vec S_{1,j}$ is imposed. 
Focusing on the vicinity of the upper critical field and applying an 
effective field theory approach, we show the
possibility of two interesting long-range-ordered states: 
for a certain high-magnetic-field area, a vector chirality 
$\langle{\cal V}_{l,j}^z\rangle=
\langle (\vec S_{l,j}\times \vec S_{l+1,j})^z \rangle$ 
or an inhomogeneous magnetization along the rung direction 
occurs in a one-component Tomonaga-Luttinger-liquid (TLL) state.
In the chiral phase, the $Z_2$ rung-parity symmetry 
$S_{l,j}^\alpha\leftrightarrow S_{l+1,j}^\alpha$, by which 
${\cal V}_{l,j}^\alpha$ changes its sign, is spontaneously
broken, while the inhomogeneous magnetization in another phase
breaks the one-site translational symmetry for the rung 
as well as the rung-parity one.   
We also predict that a two-component TLL 
emerges and all the symmetries are preserved in the 
strong-rung-coupling regime.   
Recently a spin tube material $\rm[(CuCl_2tachH)_3Cl]Cl_2$ 
(Ref.~\onlinecite{Nojiri}) has been synthesized 
and its magnetic properties could be described 
by a three-leg frustrated spin-tube model.~\cite{Mila2,Mila3,O-Y} This
also promotes the motivation of studying the spin tube~(\ref{tube}).

Existing results of the model~(\ref{tube}) are summarized here. 
In the $S=\frac{1}{2}$ case, 
the zero-field ground states are gapped and 
doubly degenerate with spontaneously breaking 
the one-site translational symmetry along the chain, at least when 
$J_\perp \agt 0.5 J$.~\cite{Ka-Ta} In addition, 
a semi-quantitative ground-state phase diagram in the 
$J_\perp$-$H$ plane ($J_\perp>0$), which only shows gapless and gapful
regimes, is constructed 
in Ref.~\onlinecite{Cab}; there exists an
intermediate magnetization plateau with $M=\langle S_{l,j}^z\rangle=1/6$.   
In the case of $S={\rm integer}$ and $H=0$, the system 
is predicted to be always gapful and to conserve all
symmetries.~\cite{MS05NLSM}

\begin{figure}[t]
\scalebox{0.8}{\includegraphics[width=\linewidth]{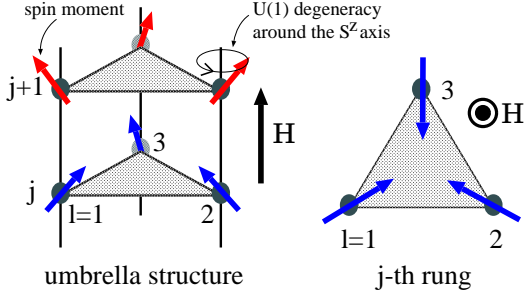}}
\caption{\label{classicalGS}Classical ground state of 
the spin tube (\ref{tube}).}
\end{figure}

Before analyzing the quantum spin tube~(\ref{tube}), 
to discuss its classical version is instructive.  
The classical ground state is 
an umbrella structure as in Fig.~\ref{classicalGS}. In this state, 
symmetries of the U(1) spin rotation around the spin z axis, 
one-site translations, and parity transformations 
along both the chain and the rung directions are all broken.   
Consequently, the system exhibits a finite vector chirality
$\langle{\cal V}_{l,j}^z\rangle
=\frac{\sqrt{3}}{2}(1-\frac{H^2}{S^2(4J+3J_\perp)^2})$. From this result, 
the vector chiral order is expected to exist even 
in the quantum version. 
However, since generally quantum fluctuation is quite strong 
in one dimension and tends to destroy any ordering, 
it is nontrivial whether or not the chiral order
remains and broken symmetries are restored in the model~(\ref{tube}).

\section{Effective theory}
\begin{figure}[t]
\scalebox{0.8}{\includegraphics[width=\linewidth]{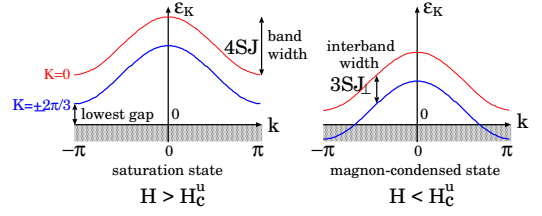}}
\caption{\label{mag_band}Magnon bands in Eq.~(\ref{one_magnon}).}
\end{figure}

Here we construct the effective theory for the quantum spin
tube~(\ref{tube}) in a high magnetic field. 
Let us begin with the fully polarized state 
with $M=S$. 
For the state, the energy dispersion of one magnon 
with $\Delta S^z=-1$ is exactly calculated as 
\begin{eqnarray}
\label{one_magnon}
\epsilon_K(k)= H-2S(J+J_\perp)+2SJ\cos k+2SJ_\perp\cos K,
\end{eqnarray}
where $K$ ($=0,\pm \frac{2\pi}{3}$) is the wave number for the rung 
and that for the chain, $k$, is in $|k|<\pi$. 
The lowest bands $\epsilon_{\pm \frac{2\pi}{3}}$ are 
always degenerate due to the rung-parity
symmetry, transformation of which induces $K\to -K$. As we explain in
Fig.~\ref{mag_band}, when $H$ becomes lower than the upper (lower)
critical value $H_c^u=4SJ+3SJ_\perp$ ($H_c^l=3SJ_\perp$), 
magnons of the lowest bands begin to condense (are fully condensed). 
Moreover, as $H<H_{c}'=4SJ$, magnons in
the remaining band $\epsilon_0$ are also condensed.

Supposing that multimagnon bound states are absent or 
their excitation energies are higher than those of one-magnon states
(this is highly expected in antiferromagnetic systems and we have 
numerically verified it near the saturation), 
we may 
describe the low-energy physics around $H\sim H_c^u$ using one-magnon
excitations. A suitable method for such a description is spin-wave
theory ($1/S$ expansion).
It makes spins bosonize as  
\begin{eqnarray}
\label{HP}
S_{l,j}^z=S-n_{l,j},&& S_{l,j}^- =b_{l,j}^\dag\sqrt{2S-n_{l,j}},
\end{eqnarray}
where $b_{l,j}$ is the magnon annihilation operator, and 
$n_{l,j}=b^\dag_{l,j}b_{l,j}$ denotes the magnon number.
Substituting  Eq.~(\ref{HP}) in the model~(\ref{tube}) and introducing 
the Fourier transformation of $b_{l,j}$ for the rung as  
\begin{eqnarray}
\label{Fourier}
b_{l,j}=\frac{1}{\sqrt{3}}\sum_{K=0,\pm 2\pi/3} e^{iK l}{\tilde b}_{K,j}, 
\end{eqnarray}
we obtain the bosonic spin-wave Hamiltonian. 
As expected, the bilinear part of $\tilde b_{K,j}$ reproduces 
the free-spin-wave dispersion $\epsilon_K(k)$. 
In order to study the low-energy and long-distance 
properties of the spin tube, we further introduce 
continuous boson fields $\Psi_q$ as follows:  
\begin{eqnarray}
\label{continuous_field}
{\tilde b}_{0,j} 
\to (-1)^j\sqrt{a_0}\Psi_0(x),\hspace{0.2cm}
{\tilde b}_{\pm\frac{2\pi}{3},j} 
\to (-1)^j\sqrt{a_0}\Psi_\pm(x),
\end{eqnarray}
where $a_0$ is the lattice spacing, and $x=ja_0$. 
Using these and taking into account 
the magnon interaction terms up to the lowest order of the $1/S$ expansion, 
we arrive in the following effective Hamiltonian, 
\begin{eqnarray}
\label{effH}
{\cal H}_{\rm eff} &&= \int dx \sum_{q=0,+,-}
\left[\frac{1}{2m_q}\partial_x\Psi_q^\dag\partial_x\Psi_q 
-\mu_q \rho_q\right] \nonumber\\
&&  +g_0\rho_0^2 + g_1(\rho_+ + \rho_-)^2 
+f_0 \rho_0(\rho_+ + \rho_-) +f_1 \rho_+\rho_-\nonumber\\
&&  +\lambda_0(\Psi_0^2\Psi_+^\dag\Psi_-^\dag+{\rm h.c.})\nonumber\\
&&+\lambda_1(\Psi_+^2\Psi_0^\dag\Psi_-^\dag
+\Psi_-^2\Psi_0^\dag\Psi_+^\dag+{\rm h.c.})+\cdots,
\end{eqnarray}
where $\rho_q=\Psi_q^\dag\Psi_q$ is the magnon-density field. 
(This Hamiltonian can also be derived via the path-integral 
approach.~\cite{K-V}) 
The first two terms 
correspond to the free-spin-wave part, 
and if the chemical potential $\mu_q$ is positive, 
the magnon $\Psi_q$ is condensed.~\cite{Note1}
We set $\mu_0=4SJ-H$ and $\mu_\pm=\mu=S(4J+3J_\perp)-H$ 
so that $H_c^u$ and $H_c'$ are fixed. 
Other parameters in Eq.~(\ref{effH}) are evaluated as
$1/m_q=2 SJ a_0^2$ ($m_q=m$), $g_0=2J a_0/3$, $g_1=(4J +3J_\perp)a_0/6$, 
$f_0=8J a_0/3$, $f_1=4J a_0/3$, $\lambda_0=(8J-3J_\perp)a_0/6$, and 
$\lambda_1=(16J-3J_\perp)a_0/12$. 
These values would be somewhat changed due to high-energy modes, 
the curvature of the dispersion, higher-order interactions, and
the hard-core property of magnons neglected in the spin-wave theory. 


\section{Lowest-band-magnon condensed state}
Based on the effective theory~(\ref{effH}), 
we investigate the spin tube near saturation. 
In this section, we consider the lowest-magnon-condensed case, 
where $\mu>0$, $\mu_0<0$, and ${\rm max}[H_c^l,H_c']<H<H_c^u$. 
For this case, the low-energy physics must be governed by two condensed
fields $\Psi_\pm$. The effective theory is derived by integrating
out the massive magnon $\Psi_0$ via the cumulant expansion 
in terms of the free-spin-wave part of $\Psi_0$ in the partition
function. The main effect of the $\Psi_0$ sector is that 
an {\it attractive} interaction between 
$\rho_+$ and $\rho_-$ originates from the second cumulant of 
the $\lambda_0$ term. 
As a result, the coupling constant $f_1$ is changed as
\begin{eqnarray}
\label{eff_coupling_cumulant}
f_1&\to&{\tilde f}_1=f_1-C
\frac{\lambda_0^2a_0^{-2}}{\sqrt{m|\mu_0|^3}}, 
\end{eqnarray}
where $C$ is a {\em positive} dimensionless constant of $O(1)$.  
Here, we have approximated the Matsubara Green's function 
$\langle T_\tau \Psi_0(x,\tau)\Psi_0^\dag(0,0)\rangle$ 
as $1/a_0$ (zero) when $|x|$ and $Ja_0\tau$ are smaller (larger) than
the correlation length $(m|\mu_0|)^{-1/2}$ [$\tau$ : imaginary time], 
and assumed that $(m|\mu_0|)^{-1/2}$ is 
at most $O(a_0)$.~\cite{Note_mu} 
For the resultant Hamiltonian 
${\cal H}_{\rm eff}'[\Psi_\pm]$, the Haldane's harmonic-fluid approach 
(i.e., bosonization) (Refs.~\onlinecite{Hal,Gia,Caza}) could be applicable. 
Using the bosonization formulas 
$\rho_\pm(x)\approx \{\bar\rho_\pm +\partial_x\phi_\pm/\pi \}
\sum_{n=-\infty}^\infty e^{i2n(\phi_\pm-\pi\bar\rho_\pm x)}$
and $\Psi_\pm^\dag \sim \{\bar\rho_\pm+\partial_x\phi_\pm/\pi\}^{1/2}
\sum_{n=-\infty}^\infty e^{i2n(\phi_\pm-\pi\bar\rho_\pm x)}
e^{-i\theta_\pm}$, 
where $\bar\rho_\pm=\langle \rho_\pm\rangle$, 
we obtain a bosonized Hamiltonian of the phase fields
$(\phi_\pm,\theta_\pm)$. Introducing the new fields 
$\phi_{s,a}=(\phi_+\pm\phi_-)/\sqrt{2}$ and 
$\theta_{s,a}=(\theta_+\pm\theta_-)/\sqrt{2}$ further, 
we can represent the phase-field Hamiltonian as 
\begin{eqnarray}
\label{bosonizedH}
{\cal H}_{[\phi,\theta]} &=& \int dx \sum_{q=s,a}
\frac{v_q}{2\pi}\left[
K_q(\partial_x\theta_q)^2+K_q^{-1}(\partial_x\phi_q)^2\right]
\nonumber\\
&&+g_{\phi} \cos(2\sqrt{2}\phi_a)
+g_{\theta}\cos(3\sqrt{2}\theta_a)+\cdots,
\end{eqnarray}
where we have assumed $\bar\rho_+=\bar\rho_-=\bar\rho$ (see below) 
and dropped terms with spatially oscillating factors $e^{i2n\pi\bar\rho x}$. 
The $g_\phi$ and $g_\theta$ terms for example 
originate from $\rho_+\rho_-$ and the third cumulant, respectively. 
Unfortunately, the values of $g_{\phi,\theta}$ cannot be evaluated
quantitatively within the present approach. 
In the phase-field picture, 
a spin rotation around the $S^z$ axis 
$S_{l,j}^+\to e^{i\gamma}S_{l,j}^+$, the one-site translation 
along the chain $S_{l,j}^\alpha\to S_{l,j+1}^\alpha$, 
that along the rung $S_{l,j}^\alpha\to S_{l+1,j}^\alpha$, 
and the site-parity transformation along the chain 
$S_{l,j}^\alpha \to S_{l,-j}^\alpha$ 
are, respectively, expressed as $\theta_\pm\to\theta_\pm+\gamma$, 
$(\phi_\pm(x),\theta_\pm(x))\to
(\phi_\pm(x+a_0)-\pi\bar\rho_\pm a_0,\theta_\pm(x+a_0)-\pi)$,  
$\theta_\pm\to\theta_\pm \pm2\pi/3$, and 
$(\phi_\pm(x),\theta_\pm(x))\to(-\phi_\pm(-x),\theta_\pm(-x))$.
Furthermore, the rung-parity transformation 
$S_{1,j}^\alpha\leftrightarrow S_{3,j}^\alpha$ 
may be realized by $\bar\rho_+=\bar\rho_-$
and $(\phi_\pm,\theta_\pm)\to(\phi_\mp,\theta_\mp)$.  
Owing to these symmetries, in all vertex operators without oscillating
factors, only $\cos[2n(\phi_+-\phi_-)]$ 
and $\cos[3n(\theta_+-\theta_-)]$ are allowed to exist
in Eq.~(\ref{bosonizedH}). 
The most relevant $n=1$ terms indeed appear in Eq.~(\ref{bosonizedH}).

The bosonization approach for ${\cal H}_{\rm eff}'$ evaluates the
velocity $v_a$ as $v_a \approx(-{\tilde f}_1\bar\rho/m)^{1/2}$. 
Therefore, if $\tilde f_1>0$, then $v_a$ becomes imaginary 
and it means that the bosonization is invalid.
To understand the physical meaning of this
instability,~\cite{Note_instability} we should consider 
the magnon-density part in ${\cal H}_{\rm eff}'$ and then define 
the following Ginzburg-Landau (GL) potential:
\begin{eqnarray}
\label{GLpotential}
{\cal F}
&=& g_1(\rho_+ +\rho_-)^2 +{\tilde f}_1 \rho_+\rho_-
-\mu(\rho_+ +\rho_-). 
\end{eqnarray}
It is clear that as $\tilde f_1>0$, the potential is minimized by
imposing $\rho_+\neq \rho_-$. 
Moreover, it is found that
\begin{eqnarray}
\label{chiral_mag_density}
\rho_+ -\rho_- \propto 
\tilde b_{\frac{2\pi}{3},j}^\dag\tilde b_{\frac{2\pi}{3},j}
-\tilde b_{-\frac{2\pi}{3},j}^\dag\tilde b_{-\frac{2\pi}{3},j}
\sim  \sum_{l=1}^3{\cal V}_{l,j}^z.
\end{eqnarray}
We thus conclude that for $\tilde f_1>0$, a finite long-range vector
chiral order $\langle {\cal V}_{l,j}^z\rangle$ exists, 
and the rung-parity symmetry is spontaneously broken.
For $J_\perp\ll J$ (i.e., $|\mu_0|/J\ll 1$) or $J_\perp\gg J$ 
[i.e., $\lambda_0\sim -O(J_\perp)$], $\tilde f_1<0$ generally
holds,~\cite{Note_strong,Note_mu} 
while for $J_\perp \sim O(J)$ (i.e., $\lambda_0\sim 0$), 
when $H$ becomes closer to $H_c^u$, 
$\tilde f_1$ increases and tends to be positive. 
Consequently, the chiral phase is present in an 
intermediate-rung-coupling regime. Supposing that $\rho_+ > \rho_-$
holds in the chiral phase, 
we can speculate that the $\Psi_-$ mode constructs a massive
spectrum, whereas the $\Psi_+$ part provides a TLL state.~\cite{K-V} 
Namely, the coexistence of the chiral order and the TLL is predicted. 
The presence of the TLL is also supported by the previous study in 
Ref.~\onlinecite{Cab}. If $H\sim H_c^u$, the TLL parameter 
would be close to the universal value 1. 
The correlation function of the chirality might exhibit 
a power decay: $\langle {\cal V}_{l,j}^z{\cal V}_{l,0}^z\rangle \approx
\langle {\cal V}_{l,j}^z\rangle^2 -{\rm const}/j^2+\cdots$ at 
$j\to\infty$.~\cite{K-V}

Let us now discuss the case of $\tilde f_1<0$, where 
$\bar\rho_+=\bar\rho_-$ is restored and the bosonization is available. 
The $\phi_s$ sector in Eq.~(\ref{bosonizedH}) yields a TLL, 
which is strongly stabilized by symmetries, 
while the low-energy physics of the $\phi_a$ sector depends on whether
$\cos(2\sqrt{2}\phi_a)$ and $\cos(3\sqrt{2}\theta_a)$
are relevant or not: the scaling dimensions of these two are $2K_a$
and $9/(2K_a)$, respectively. 
The Hamiltonian ${\cal H}_{\rm eff}'$ leads to 
$K_a\propto (-\bar\rho/\tilde f_1)^{1/2}$. 
Therefore, when $\tilde f_1\sim 0$ and $\bar\rho$ is large enough, 
$K_a$ is always much larger than 1. 
At this case, $\cos(3\sqrt{2}\theta_a)$ and $\cos(2\sqrt{2}\phi_a)$ 
are respectively highly relevant and irrelevant, and then the $\phi_a$
sector obtains a massive spectrum. If $g_\theta>0$ ($<0$), the phase
field $\theta_a$ is pinned on lines $\theta_a=\sqrt{2}(2n+1)\pi/6$ 
($\sqrt{2}n\pi/3$) in the $\theta_+$-$\theta_-$ plane. 
Among these lines, only six lines intersect the physically 
relevant ``Brillouin'' zone, $-\pi<\theta_+\leq \pi$ and 
$-\pi\leq\theta_-< \pi$. This result implies that 
the ground states possess the sixfold degeneracy. 
To investigate the physical meaning of locking $\theta_a$ and the
ground-state degeneracy, let us focus on the magnetization per site. 
The bosonization represents it as
\begin{eqnarray} 
\label{inhomo_mag}
\langle S_{l,j}^z\rangle \approx 
M-\frac{2}{3} \bar\rho a_0
\Big\langle\cos\Big(\sqrt{2}\theta_a+\frac{4}{3}\pi l\Big)\Big\rangle 
+\cdots.
\end{eqnarray}
One can see that the second term in Eq.~(\ref{inhomo_mag})
causes a down-down-up magnetization structure in the case of
$g_\theta>0$, while for $g_\theta<0$ an up-up-down structure occurs: 
for instance, if $\theta_a$ is locked to zero for $g_\theta<0$, 
$\langle S_{1,j}^z\rangle=\langle S_{2,j}^z\rangle=M+\delta$ and 
$\langle S_{3,j}^z\rangle=M-2\delta$ 
[$\delta\propto\langle\cos(\sqrt{2}\theta_a)\rangle$]. 
We thus conclude that an inhomogeneous magnetization for the rung 
is induced by pinning $\theta_a$.
Obviously, the parity and translational symmetries for the rung direction
are spontaneously broken in this state. 
Three of the sixfold degenerated
states are indeed explained by this inhomogeneous distribution. 
The meaning of the remaining twofold degeneracy is unknown.~\cite{Note_dege} 
Remarkably, the inhomogeneously magnetized phase is not at all expected 
from the classical tube system (see Fig.~\ref{classicalGS}). 
We note that this inhomogeneous distribution might slightly be modified 
if $\cos(3n\sqrt{2}\theta_a)$ with $n\geq 2$ are also 
relevant.~\cite{Note-otherterms}
From the predictions of the chiral order for $\tilde{f}_1>0$ and the
inhomogeneous phase under the condition $\tilde{f}_1< 0$ and 
$|\tilde{f}_1|\sim 0$, the boundary $\tilde f_1=0$ is expected to 
be a first-order transition.

When $-\tilde f_1/\bar\rho$ increases so that 
$K_a<9/4$,  
$\cos(3\sqrt{2}\theta_a)$ becomes irrelevant and the low-energy physics
of the $\phi_a$ sector is described by a Gaussian model. 
This transition must be of a Beresinskii-Kosterlitz-Thouless (BKT)
type.~\cite{Gogo} After the transition, the system is in a two-component 
TLL phase with all symmetries enjoying. 
If $-\tilde f_1/\bar\rho$ is further increased due to 
the growth of $J_\perp$ or the decrease of $\bar\rho$, 
$\cos(2\sqrt{2}\phi_a)$ seems to become relevant. 
However, the exact results for the integrable Bose gas~\cite{Caza2}
imply that in a one-dimensional Bose system with a short-range repulsive 
interaction, the TLL parameter is not usually smaller than 1 even when 
the interaction becomes extremely strong. 
The two-component TLL is hence expected to continue 
even when $J_\perp\gg J$ or $\bar\rho$ is small
(see the Endnote~\onlinecite{Note_strong}). 
The prediction of the two-component TLL in the
strong-rung-coupling regime is in agreement with a previous 
study applying the strong-rung-coupling approach to the 
$S=\frac{1}{2}$ tube.~\cite{Cit}

\section{Three-band-magnon condensed state}
Here, we consider the case where 
all three kinds of magnons $\Psi_{+,-,0}$ are condensed. 
This situation could be realized under the condition of 
$\mu>0$, $\mu_0>0$, $H_c^l<H<H_c'$, and $J_\perp<4J/3$. 
This means that the three-band-magnon condensed state is allowed to
exist only in the weak-rung-coupling regime. 
Like Eq.~(\ref{GLpotential}), let us
introduce the GL potential for the present case as follows:   
\begin{eqnarray}
\label{GLpotential_v2}
{\cal G}
&=& g_0\rho_0^2 + g_1(\rho_+ +\rho_-)^2
+f_0 \rho_0(\rho_+ +\rho_-)
\nonumber\\
&&+ f_1 \rho_+\rho_- 
-\mu_0\rho_0 - \mu(\rho_+ +\rho_-).
\end{eqnarray}
To find the stable magnon-density profile 
$(\rho_0,\rho_+,\rho_-)$, the Hessian matrix 
$H_{i,j}=[\frac{\partial^2 {\cal G}}{\partial\rho_i\partial\rho_j}]$ is
useful. At the local minimum point $(\rho_0,\bar\rho,\bar\rho)$ satisfying 
$\partial{\cal G}/\partial\rho_j=0$, the eigenvalues of $H_{i,j}$ are 
$-4J/3$, $C_1$, and $C_2$ ($-4J/3<C_1<0$ and $C_2>0$).
The corresponding eigenvectors are 
$(\delta\rho_0,\delta\rho_+,\delta\rho_-)\propto(0,1,-1)$, $(-C_3,1,1)$, 
and $(C_3,1,1)$, where $C_3>0$. 
The negative eigenvalue $-4J/3$ and its 
eigenvector indicate that the ground state takes $\rho_+ -\rho_-\neq 0$. 
Moreover, a positive eigenvalue $C_2$ implies the existence of the TLL.  
We therefore predict that the chiral order ($\rho_+\neq\rho_-$) 
and a one-component TLL state still remain 
when the system moves from the lowest-magnon-condensed regime 
to all-magnon-condensed one.~\cite{note_FMrung,pre}    
At the boundary between these two regime, one might observe a weak
singularity such as a magnetization cusp.


\section{Summary and discussions}
We have studied the three-leg frustrated 
spin tube~(\ref{tube}) near the upper critical field. It has been 
predicted that the vector chiral order or the inhomogeneously magnetized 
order emerges in the magnetic-field-driven TLL phase 
in the intermediate-rung-coupling regime. It is remarkable that in these
two phases, the TLL criticality (massless modes) and the spontaneous
breakdown of {\it discrete} parity or translational symmetries for
the rung direction coexist. 
We have also shown that when the rung coupling becomes strong enough, 
the inhomogeneous phase vanishes and instead the two-component TLL 
occurs with preserving all the symmetries.

\begin{figure}[t]
\scalebox{0.8}{\includegraphics[width=\linewidth]{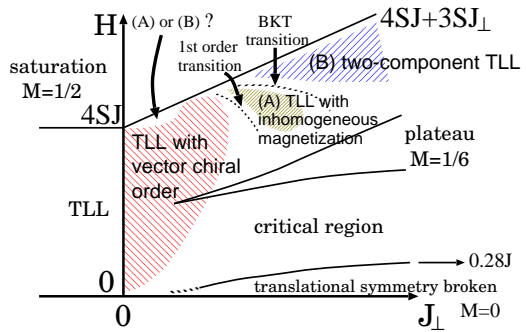}}
\caption{\label{phases}Schematic ground-state 
phase diagram of the $S=\frac{1}{2}$ spin tube~(\ref{tube}). 
The area away from the saturation is discussed 
elsewhere (Ref.~\onlinecite{pre}). 
See the Endnotes~\onlinecite{Note_mu} and \onlinecite{Note_strong}.}
\end{figure}

Combining our results and the existent ones,~\cite{Ka-Ta,Cab} 
we can draw the ground-state phase diagram for the $S=\frac{1}{2}$ tube 
as in Fig.~\ref{phases}. 
The global phase structure near the saturation 
would common to all the cases with arbitrary $S$, as far as 
$S \alt O(1)$. Although in general the spin-wave approach used 
in this paper is not very reliable for small-$S$ cases, we believe that
it is valid if we consider the region where $M$ is sufficiently close 
to the saturation value: in such a region, multimagnon scattering
processes are expected to be negligible.    
When $J_\perp$ is changed from $+0$ to $+\infty$ with 
$M$ fixed near the saturation, 
the following scenario is expected: 
TLL plus chirality
$\to$[first-order transition]$\to$ TLL plus inhomogeneous magnetization  
$\to$[BKT transition]$\to$ two-component TLL. 

We finally note that
the predicted first-order and BKT transitions could not be detected 
by observing the magnetization $M$ because 
$H$ couples to $\partial_x\phi_s$ and $\rho_+ +\rho_-$, 
but it does not directly interact $\phi_a$ and $\rho_+ -\rho_-$.
A specific-heat measurement would be efficient in the detection.

\begin{acknowledgments}
This work is supported by a Grant-in-Aid for Scientific 
Research (B) (No. 17340100) from the Ministry of Education, 
Culture, Sports, Science and Technology of Japan.
\end{acknowledgments}





\begin{thebibliography}{}
\bibitem{Note_mu}As $|\mu_0|/J$ approaches zero, 
	the spatial range of the interaction between $\Psi_+$ and $\Psi_-$
	becomes larger. For such a case, the bosonization approach in
	this paper is less valid.
\bibitem{Note_instability}When in Ref.~\onlinecite{MS05NLSM} 
	we discussed the possibility of the
	two-component TLL in frustrated integer-spin tubes near the
	lower critical field, this type instability induced by the
	density-density interaction was not taken into account.
\bibitem{Note_strong}Note that in the extremely strong-rung-coupling
	area, the spin-wave approach becomes less reliable because some
	coupling constants are much larger than the spin-wave band
	width.
\bibitem{Note_dege}The remaining degeneracy might be related with
a nonlocal symmetry breaking.
\bibitem{Note-otherterms}The modification of the magnetization would
	strongly depend on coupling constants of
	$\cos(3n\sqrt{2}\theta_a)$. However, we cannot quantitatively 
	calculate them within the present approach.
\bibitem{note_FMrung}If a similar analysis using the Hessian is applied
	to the ferromagnetic- (FM-) rung case, the chiral order is
	predicted to exist even in $J_\perp<0$. 
	However, the expectation is doubtful because magnon bound states
	would be important in the FM-rung regime.
\end{thebibliography}
\end{document}